\documentclass[referee]{raa}            

\usepackage{graphicx}             
\usepackage{natbib}
\usepackage{amssymb,amsmath}
\usepackage{longtable}
\usepackage{booktabs}
\bibpunct{(}{)}{;}{a}{}{,}

\usepackage[a4paper=true,pagebackref=true]{hyperref}
\hypersetup{colorlinks = true, linkcolor = green, anchorcolor = red, citecolor = blue, filecolor = red, pagecolor = red, urlcolor = red}

\begin{document}

   \title{Absolute physical parameters of three poorly studied detached eclipsing binaries
}

   \volnopage{Vol.0 (20xx) No.0, 000--000}      
   \setcounter{page}{1}                         
   \author{D.-Y. Yang
      \inst{1,2,3}
   \and L.-F. Li
      \inst{1,2,4,5}
   \and Q.-W. Han
      \inst{6}
   }

   \institute{Yunnan
Observatories, Chinese Academy of Sciences, 650216, Kunming,
P.R. China; {\it llf@ynao.ac.cn}\\
        \and
             Key Laboratory for the Structure and Evolution of
Celestial Objects, Chinese Academy of Sciences, 650216,
Kunming, P.R. China\\
	\and
University of the Chinese Academy of Sciences, Beijing, 100049, P.R. China\\
\and 
Center for Astronomical Mega-Science, Chinese Academy of Sciences, 20A
Datun Road, Beijing, 100012, P.R. China\\
\and
Key Lab of Optical Astronomy, National Astronomical Observatories, Chinese Academy of Sciences, 20A Datun Road, Chaoyang, Beijing, 100012, P.R. China\\
\and
School of Physics and Electronics, Qiannan Normal University for Nationalities, Duyun 558000, PR China\\
\vs \no
}

\abstract{ 
The photometric and spectroscopic data for three double-lined detached eclipsing binaries were collected from the photometric and spectral surveys. The light and radial velocity curves of each binary system were simultaneously analyzed by using Wilson-Devinney (WD) code, and the absolute physical and orbital parameters of these binaries were derived.  The masses of both components of ASASSN-V J063123.82+192341.9 were found to be  $M_1 = 1.088 \pm 0.016$ and $M_2 = 0.883 \pm 0.016\ M_{\odot}$;  and those of ASAS J011416+0426.4 were determined to be $M_1 = 0.934 \pm 0.046$ and $M_2 = 0.754 \pm 0.043 M_{\odot}$; those of MW Aur were derived to be $M_1 = 2.052 \pm 0.196$ and $M_2 = 1.939 \pm 0.193\ M_{\odot}$. At last, the evolutionary status of these detached binaries was discussed based on their absolute parameters and the theoretical stellar models.
}

   \authorrunning{D.-Y. Yang et al. }            
   \titlerunning{Absolute parameters of three detached binaries}  

   \maketitle

%
%
\section{Introduction}           

\label{sect:intro}

Various questions about stellar structure and evolution remain to be answered and also require precise and detailed information on the physical parameters, such as mass, radius, luminosity and effective temperature \citep{2010Ap&SS.328...51C, 2015PASA...32...28E}. Meanwhile, the mass of a star is the most critical parameter, which determines the way it evolves and what is left over after its death \citep{2015NewA...41...42B}. By far, the accurate method to derive the masses of the stars is through the photometric and spectroscopic observations of eclipsing binaries. Current observational techniques can produce masses to a precision of better than $3\%$, accurate enough to provide strong constraint for stellar models with inadequate physics to be rejected \citep{2009ApJ...700.1349T}. Particularly, detached eclipsing binaries contain two components, which have not filled their Roche lobe yet (i.e. they have not been contaminated by the mass transfer between them) and thus evolve as single stars \citep{2015NewA...41...42B}. Therefore, the two components of detached eclipsing binaries with well determined parameters can provide a stringent test for the stellar evolutionary models. 

Short-period binaries usually have an orbital eccentricity close to zero due to the strong tidal friction, and some long-period ones sometimes have an eccentric orbit. In addition, double-lined spectroscopic binaries, especially those with a low mass ratio, are rare, since the secondary star is too faint to be spectroscopically observed at present \citep{2019AJ....158..198F}. However, the double-lined spectroscopic binaries can give us the opportunity to obtain their orbital parameters, and to determine their precise masses together with other physical parameters \citep{2019A&A...621A..93S}. Moreover, detached binaries are the progenitors of many peculiar objects \citep[e.g. W UMa binaries and blue stragglers,][]{1982A&A...109...17V, 2020MNRAS.492.2731J}, and detailed observation investigation of detached binaries are also important to understand the formation of these peculiar objects.

In this work, we selected three detached eclipsing binaries from the Large Sky Area Multi-Object Fiber Spectroscopic Telescope Media-Resolution Spectra Survey\citep[LAMOST-MRS,][]{1996ApOpt..35.5155W, 1997ASSL..212...67C, 1999oaaf.conf....1Z, 2012RAA....12..723Z, 2006ChJAA...6..265Z, 2015RAA....15.1095L}, and determined their  mass ratio based on the analysis of their radial velocity curves. The photometric observations in $V$ pass-band for these binaries were collected from the All-Sky Automated Survey\citep[ASAS,][]{1997AcA....47..467P}, the All-Sky Automated Survey for supernova \citep[ASAS-SN,][]{2014ApJ...788...48S} and the Wide Angle Search for Planets \citep[WASP,][]{2010A&A...520L..10B}. The spectroscopic and photometric data were analyzed simultaneously by using the pyWD2015 code \citep{2020CoSka..50..535G}. In this process, we determined the accurate orbital and physical parameters for the three objects and discussed their evolution. The basic information of the three objects are listed in Table~\ref{table:table1}.

This paper is organized as follows. In Sect~\ref{sec:two}, the origin of the photometric and spectroscopic data used in this work are stated. In Sect~\ref{sec:three}, the method used for our study and the results of the derived accurate parameters are described. Our discussions and conclusions are shown in Sect~\ref{sec:five}.

\begin{table}[h]
\caption{Basic data for ASASSN-V J063123.82+192341.9, ASAS J011416+0426.4 and MW Aur} 
\label{table:table1} 
\resizebox{\textwidth}{!}{
\centering
\begin{tabular}{ccccccccccccccccc}
\hline\noalign{\smallskip}
Name & $R.A.$ & $Decl.$ & $J^{(a)}$& $K^{(a)}$ & $B^{(a)}$ & $V^{(a)}$ & $T_{\rm eff}^{(b)}$& $[Fe/H]$ & $P_{\rm obs}^{(c)}$ & $Epoch^{(c)}$  \\
 & (h:m:s) & (deg:m:s) & (mag) & (mag) & (mag) & (mag) & (K)&  & (d) &(d) 
\\
\hline\noalign{\smallskip}
ASASSN-V J063123.82 $+$192341.9 & 06:31:23.84 & +19:23:43.26 & 11.850 (0.02) & 11.486 (0.02) & 13.850 (0.10) &13.195 (0.11) & 6254 (145)& -0.024 (0.069)$^{(d)}$ & 1.3044 & 58021.0809  \\
ASAS J011416$ +$0423.4          & 01:14:15.97 & +04:26:23.56 & 9.323 (0.03)  & 8.858 (0.02)  & 11.289 (0.06)&10.608 (0.02) & 5621 (58) & -0.147(0.121)$^{(e)}$ & 2.4780 & 51889.1600   \\
MW Aur   & 06:02:29.12 & +29:20:09.76 & 10.908 (0.02)  & 10.607 (0.02) & 12.635 (0.04) & 12.088 (0.03)& 7242 (366)& -0.038 (0.144)$^{(d)}$ & 15.3405 & 54085.7000  \\

\hline\noalign{\smallskip}
\end{tabular}}
\tablecomments{0.86\textwidth}{$^{(a)}$Observed magnitudes in V, J and B band from UCAC4 Catalogue \citep{2013AJ....145...44Z}. $^{(b)}$Stellar effective temperatures from Gaia DR2 \citep{2019AJ....158...93B}. $^{(c)}$Period and epoch of primary minimum of light curve from VSX catalogue \citep{2006SASS...25...47W}.  $^{(d)}$Metallicity from the LAMOST-MRS Parameter Catalogue \citep{2020ApJ...891...23W}. $^{(e)}$Metallicity from the catalogue of the stellar parameters for EAs observed by LAMOSY \citep{2018ApJS..235....5Q}.}
\end{table}

\section{Observation data} \label{sec:two}

\subsection{Photometry data} \label{subsec:tables}
The photometric data for three detached eclipsing binaries were collected from various photometric surveys, and the orbital periods of these binaries are too long to obtain the complete light curve(s) easily. Only data points with a relatively high precision were used.

For ASASSN-V J063123.82+192341.9, a total of 111 measurements in $V$ band were collected from ASAS-SN \citep{2017PASP..129j4502K,2018MNRAS.477.3145J}, and all measurements were used in deriving the photometric solution for this eclipsing binary since they have a relatively high observational accurancy (better than 0.03 mag). For ASAS J011416+0426.4, a total of 528 observations in $V$ band were obtained from ASAS \citep{2001ASPC..246...53P, 2008AcA....58..405S}, only 469 measurements were employed to obtain a light curve solution for this object after 59 scattered data points were removed. In addition, a total of 3067 observations in $V$ band for MW Aur were obtained from WASP \citep{2010A&A...520L..10B}, but only 2723 data points were used for analysis after the scattered points and those with an error higher than 0.1 mag were removed.

\subsection{Spectroscopy data} \label{subsec:tables}

LAMOST is a special reflecting Schmidt telescope with an effective aperture of 3.6$\times$4.9m, a focal length of 20m and a field of view (FOV) of $5^\circ$, which locates at Xinglong station, Hebei Province, China \citep{1996ApOpt..35.5155W, 2012RAA....12..723Z}. The focal surface has 4000 precisely positioned fibers connected to 16 spectrographs with a distributive parallel-controllable fiber positioning system, each spectrograph equipped with a 4K $\times$ 4K Charge-Coupled Devices (CCD) for blue and red channels \citep{1998SPIE.3352..839X}. By the end of 2017, all 16 media$-$resolution spectrographs were in operation, and had prepared for a new five-year medium-resolution spectroscopic survey (started in September 2018). The MRS  operates at 4950$\AA < \lambda < 5350\AA$ (B band) and 6300$\AA < \lambda < 6800\AA$ (R band) with a spectral resolution of R $\thicksim$ 7500 \citep{liu2020lamost}. In this study, the media-resolution spectra were all collected from LAMOST-MRS.

\section{Data analysis and results} \label{sec:three}

\subsection{Radial Velocities (RVs)} \label{subsec:tables}

\begin{figure}[ht!]
\includegraphics[height=7cm,width=1.0\textwidth]{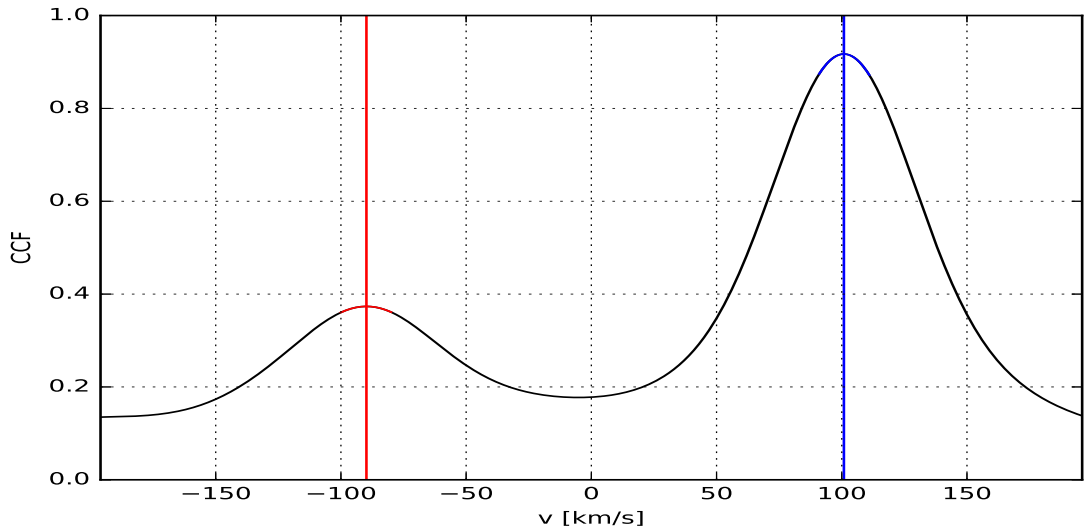}
\caption{An example of calculating RVs for ASAS J011416+0426.4, the blue and read line represent the RVs of  primary and secondary, respectively.  
\label{fig:genera1}}
\end{figure}

In the process of calculating the RVs for both components of the binary systems, we employed the RaveSpan software \citep{2013MNRAS.436..953P,2015ApJ...806...29P,2017ApJ...842..110P}, in which includes three major velocity extraction methods: cross-correlation function\citep[CCF;][]{1974A&A....31..129S,1979AJ.....84.1511T}, two dimensional cross-correlation \citep[TODCOR;][]{1994ApJ...420..806Z} and the broadening function technique \citep[BF;][]{2002AJ....124.1746R}. For media-resolution spectra, we chose the CCF pattern of the RaveSpan software to derive RVs. However, this method relies heavily on the sign-to-noise ratio (S/N) of the observed spectra, so we made use of the co-added LAMOST DR7 \footnote{\url{http://dr7.lamost.org/doc/mr-data-production-description}} B band spectra, which produced by combing the single exposure spectra with a relatively low S/N. Meanwhile, the template spectra used in our study were selected from \cite{2018A&A...618A..25A}. The values of CCF for double-lined spectrum usually present the double peaks, which represent the RVs of the primary and secondary respectively. During the eclipse, usually only one peak can be measured, which represents the RV of the mass center. We chose "4th-order polynomial" method of RaveSpan software to fit each CCF peak, then obtained RVs of each component of these binary systems. An example is shown in Figure~\ref{fig:genera1}.  The derived RVs of all objects are shown in Table~\ref{rvs}. 

\begin{table}[h]
\centering
\caption[]{The derived RVs of the primary and secondary
components for three objects.\label{rvs}}
\resizebox{\textwidth}{!}{
\begin{tabular}{ccccccccccccccccc}
\hline\noalign{\smallskip}
\multicolumn{3}{c}{ASASSN-V J063123.82+192341.9}
 &  & \multicolumn{3}{c}{ASAS J011416+0426.4}
 &  & \multicolumn{3}{c}{MW Aur} 
 \\
\cline{1-3}
\cline{5-7}
\cline{9-11}
HJD & Primary & Secondary &  & HJD & Primary & Secondary &  & HJD & Primary & Secondary\\
2400000+ & (km s$^{-1}$)
& (km s$^{-1}$) &&2400000+ & (km s$^{-1}$)
& (km s$^{-1}$)&&2400000+ & (km s$^{-1}$)
& (km s$^{-1}$)

\\
\hline\noalign{\smallskip}
   58415.35521 & -143.836(0.107)  &  97.143(0.341) && 58411.14931 & -6.456(0.092)   &  $\cdots$  & &   58056.36563 & -45.452(0.103)   &  80.823(0.111)    \\
   58482.20764 & -37.643(0.105)  &    $\cdots$     & &58416.15695 &  12.369(0.080)  &  $\cdots$  &  &58057.36355 & -91.144(0.093)   &  125.623(0.105)    \\
   58496.11979 & -129.828(0.111)  &  78.767(0.313) & &58420.14098 & -28.192(0.100)  & 70.200(0.491) & &  58060.34271 &  22.283(0.081)   &  $\cdots$          \\
   58499.12535 & -63.318(0.125)   &    $\cdots$     && 58439.05452 &  101.941(0.080) & -87.881(0.306)& & 58062.27188 &  52.277(0.189)   &  -20.419(0.216)   \\
   58528.07327 &  52.267(0.118)   &  -146.593(0.354)&&58449.04202 &  104.145(0.081) & -88.157(0.298) & & 58066.21736 &  46.864(0.471)   &  -19.718(0.351)   \\
   58542.04584 & -117.912(0.133)  &   64.975(0.401) & &58451.07500 &  57.735(0.115)  &  $\cdots$     &   &58532.02049 & -46.874(0.109)   &  91.554(0.119)    \\
                                 &   &  && 58453.03091 & -43.571(0.097)  & 95.443(0.343) &  & 58536.00938 &  21.168(0.089)   &  $\cdots$          \\
                                 &   &  & &58466.02848 &  72.548(0.097)  & -44.054(0.529)&   & &    &          \\                                                                 
       							 &   &  & &58477.96702 & -9.445(0.110)   &  $\cdots$   &&    &     &            \\                              
\hline\noalign{\smallskip}
\end{tabular}}
\end{table}

\subsection{The simultaneous solution of light and RV curves for three objects} \label{subsec:four}
The light curves in $V$ band and radial velocity curves of three eclipsing binaries were analyzed through
2015 version of WD code \citep[named pyWD2015, a “GUI wrapper for WD code,][]{1971ApJ...166..605W,1979ApJ...234.1054W,1990ApJ...356..613W,2012AJ....144...73W}, which provides a convenient interface to input parameters and run DC and LC programs for the users \citep{2020CoSka..50..535G}. 

The input fixed parameters for the DC subroutine would decide whether the appropriate solution can be obtained for a binary system or not. At first, the mean effective temperature of the primary component of each binary system was obtained from Gaia DR2 \citep{2019AJ....158...93B}. The rotation parameters ($F_{1,2}$) are defined as a ratio of the axial rotation rate to the mean orbital rate for both components of a binary system. Therefore, they were  set to be 1.0 for  ASASSN-V J063123.82+192341.9 and ASAS J011416+0426.4 since the orbital periods of two binaries are relatively short and their components should rotate synchronously with orbital motion because of tidal friction. The rotation parameters of both components of MW Aur were set to be 5.0 (given by $F = \frac{P_{\rm orb}}{P_{\rm rot}}$, where $P_{\rm orb}$ and $P_{\rm rot}$ denote orbital period and rotation period, respectively) since this binary has an elliptical orbit \citep[$P_{\rm rot}$ = 3.068,][]{2018AJ....155...39O} and a relatively long orbital period. In addition, two binary systems (ASASSN-V J063123.82+192341.9 and ASAS J011416 +0426.4) should have a convective atmosphere since their primary components have a low surface effective temperature (see Table~\ref{table:table1}), thus the bolometric albedos and the gravity darkening coefficients were taken to be 0.5 and 0.32 for both components of two binaries, respectively \citep{1969AcA....19..245R}. Since the binary system MW Aur should have a radiative atmosphere according to the effective temperature of its primary component, then the two coefficients mentioned above were all taken to be 1.0 for two components of MW Aur \citep{1924MNRAS..84..665V}. Meanwhile, we chose a logarithmic law \citep{1970AJ.....75..175K} to determine the limb darkening coefficients for both components of these eclipsing binaries [i.e. set LD1 (LD2) = $\pm2$]. At last, a simple reflection treatment (with parameters MREF = 1, NREF = 1) was chosen. The orbital eccentricities of our targets except for MW Aur were set to be 0, since the eclipsing system MW Aur exhibits an asymmetric light curve in V-band evidently. Therefore, the adjustable parameters used in WD models are as the followings: the orbital semi-major axis ($a$), the systemic velocity ($V_\gamma$), the orbital inclination ($i$), the mean surface temperature of secondary star ($T_2$), the modified surface potential of both components ($\Omega_1$ and $\Omega_2$), the mass ratio ($q$), the bandpass luminosity of primary star ($L_1$), the third light ($l_3$), the epoch of primary minimum ($T_0$) and orbital period ($P$). In addition, the orbital eccentricity ($e$), the argument of periastron ($\omega$) and the phase shift ($\phi_0$) were also set as free parameters for MW Aur.

\begin{figure}[ht]
\centering
\includegraphics[width = \textwidth]{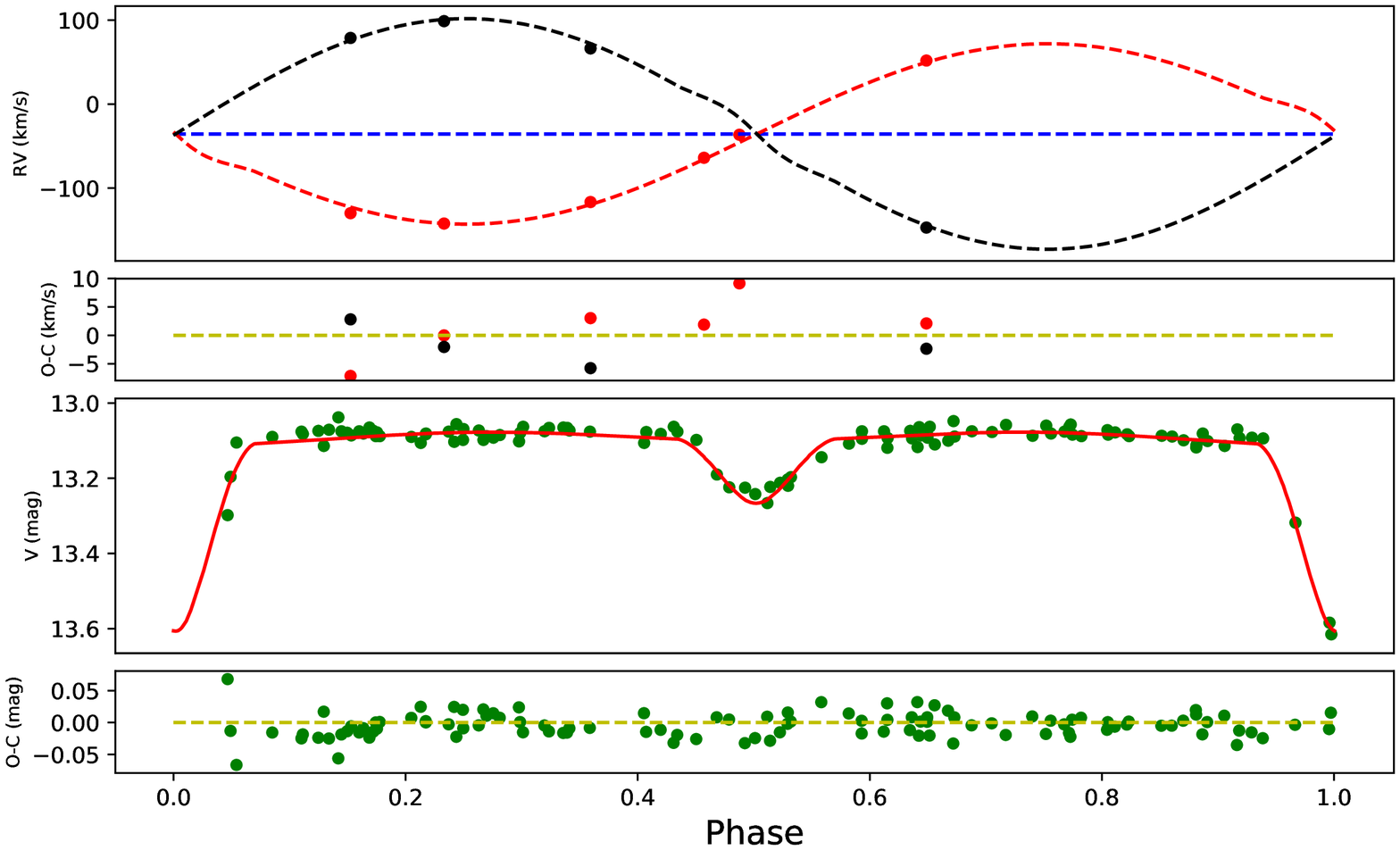}
\caption{The observed radial-velocity (top panel) and V-band light (bottom panel) curves (solid points) and the best fits carried out by WD code in dashed or solid lines for ASASSN-V J063123.82+192341.9. The fitting residuals are presented at the bottom of each panel.
\label{mode1}} 
\end{figure}

\begin{figure}[ht]
\centering
\includegraphics[width = \textwidth]{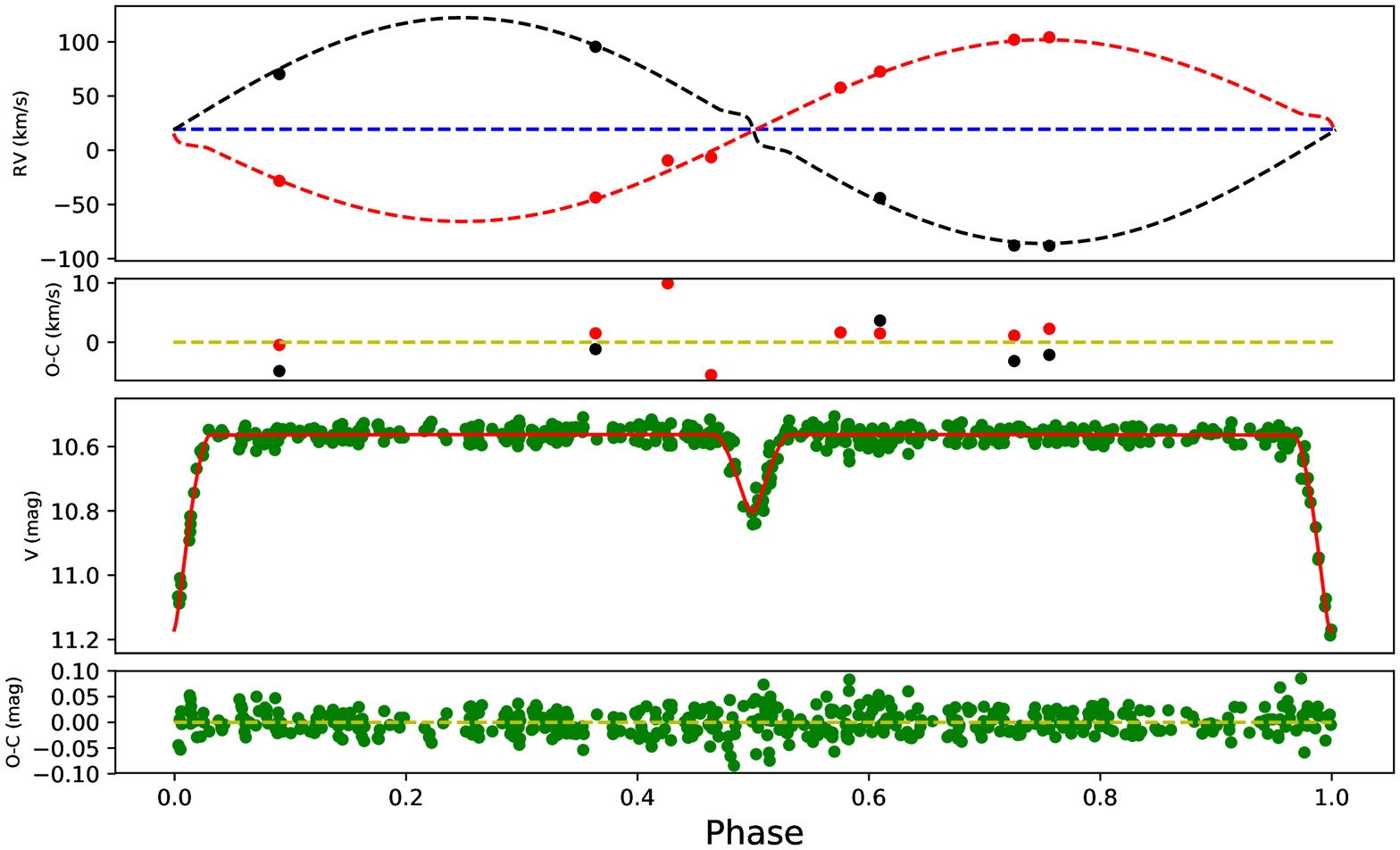}
\caption{The observed radial-velocity (top panel) and V-band light (bottom panel) curves (solid points) and the best fits carried out by WD code in dashed or solid lines for ASAS J011416+0426.4. The fitting residuals are presented at the bottom of each panel.
\label{mode2}} 
\end{figure}

\begin{figure}[ht]
\centering
\includegraphics[width=\textwidth]{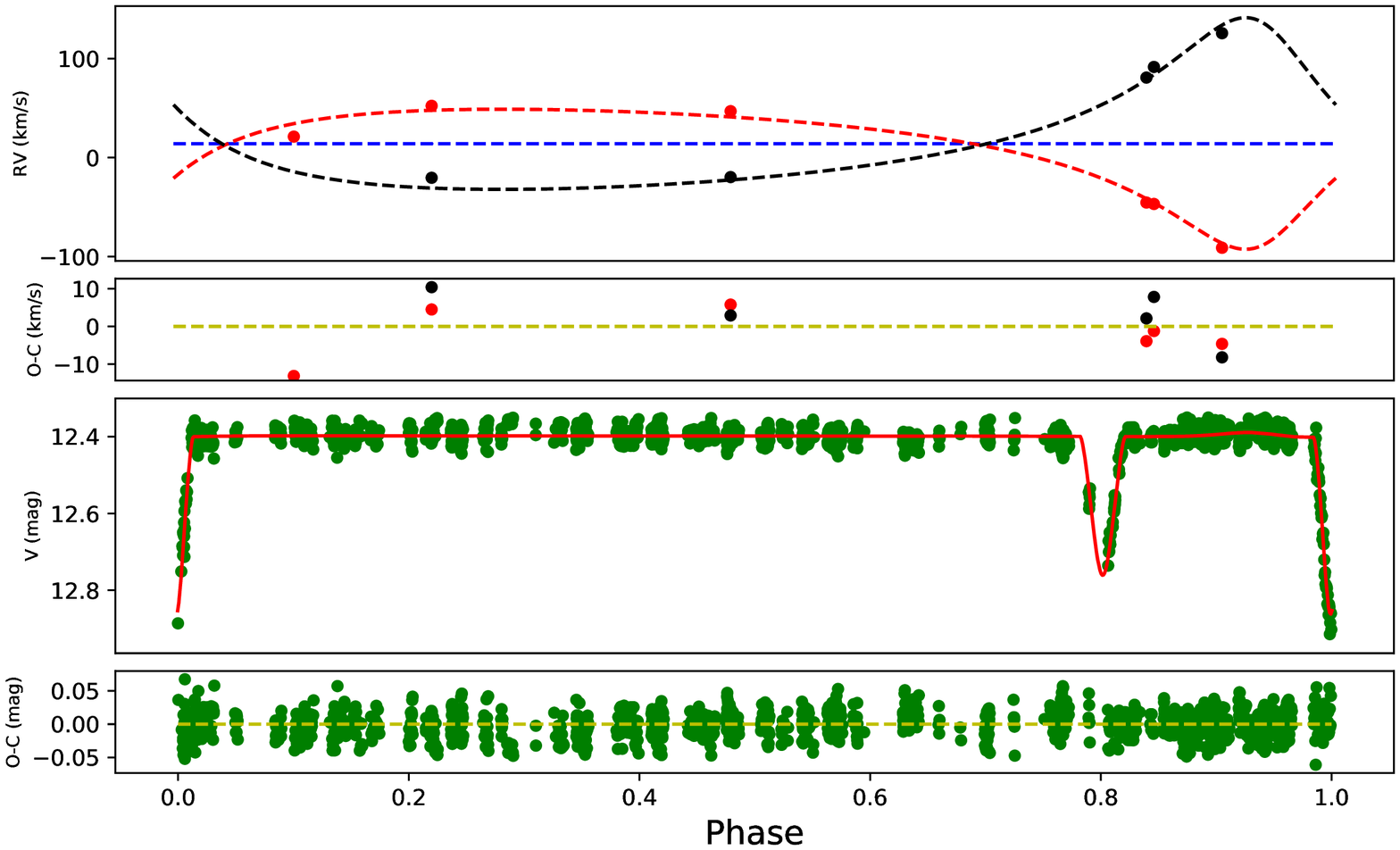}
\caption{The observed radial-velocity (top panel) and V-band light (bottom panel) curves (solid points) and the best fits carried out by WD code in dashed or solid lines for MW Aur. The fitting residuals are presented at the bottom of each panel.
\label{mode3}} 
\end{figure}

The calculation for each binary system always started at mode 2 (detached mode), then the best solutions for $V$-band light and radial velocity curves were gotten from multiple iterations using automated differential correction (DC) optimizing subroutine of the WD code until they converged. In the calculation, we found that the solutions for all three eclipsing binaries were converged at mode 2, suggesting that three eclipsing binaries all have a detached configuration at present. We also checked whether the third light ($l_3$) has influence on these solutions, and found that all cases show an unphysical value for the third light ($l_3 < 0$), and therefore we adopted $l_3 = 0$ in our final solution. The convergent solutions derived from the V-band light curves and the radial-velocity curves are presented in Table~\ref{table:table3}, and the results are shown in Figure~\ref{mode1}, Figure~\ref{mode2} and Figure~\ref{mode3}, respectively. 

\begin{table}[h]
\caption{Photometric parameters of three detached eclipsing binaries.} 
\label{table:table3} 
\resizebox{\textwidth}{!}{
\centering
\begin{tabular}{ccccccccccccccccc}
\hline\noalign{\smallskip}
Parameter & \multicolumn{2}{c}{ASASSN-V J063123.82+192341.9}
  &  & \multicolumn{2}{c}{ASAS J011416+0426.4} 
  &  & \multicolumn{2}{c}{MW Aur} 
 \\
\cline{2-3}
\cline{5-6}
\cline{8-9}
 & Primary & Secondary && Primary & Secondary & \ & Primary & Secondary\\
 \hline\noalign{\smallskip}
$Period$(d)&\multicolumn{2}{c}{1.304 (fixed)}& &\multicolumn{2}{c}{2.478 (fixed)}&&\multicolumn{2}{c}{15.304 (fixed)}\\
$T_0$(HJD-2400000)&\multicolumn{2}{c}{58021.125 (fixed)}&&\multicolumn{2}{c}{51890.389 (fixed)}&&\multicolumn{2}{c}{54085.679 (fixed)}\\
$T_{\rm eff}$(K)&6254 $\pm$ 145&5127 $\pm$ 54& &5621 $\pm$ 58&4883 $\pm$ 31& &7242 $\pm$ 366&6613 $\pm$ 34\\
$i(\rm deg)$   &  \multicolumn{2}{c}{88.759 $\pm$ 1.425}&&\multicolumn{2}{c}{88.323 $\pm$ 0.685}&&\multicolumn{2}{c}{84.739 $\pm$ 0.067}\\
$e$ &\multicolumn{2}{c}{0}&&\multicolumn{2}{c}{0}&&\multicolumn{2}{c}{0.521 $\pm$ 0.0.002}\\
$q$ &\multicolumn{2}{c}{0.811 $\pm$ 0.009}&&\multicolumn{2}{c}{0.808 $\pm $0.023}&&\multicolumn{2}{c}{0.945 $\pm$ 0.027}\\
$\Omega$ &5.081 $\pm$ 0.110&6.888 $\pm$ 0.214& &9.763 $\pm$ 0.246&11.359 $\pm$ 0.448& &19.903 $\pm$ 0.217&12.933 $\pm$ 0.263    \\
$\Omega_{\rm in}$&3.434&2.979& &3.417&2.967& &6.465&5.971   \\
$\omega$  & \multicolumn{2}{c}{0}  &    & \multicolumn{2}{c}{0} & &  \multicolumn{2}{c}{201.153 $\pm$ 0.656}\\
$V_\gamma$(km s$^{-1}$)&\multicolumn{2}{c}{-35.673 $\pm$ 0.001}& &\multicolumn{2}{c}{19.338 $\pm$ 0.725}& &\multicolumn{2}{c}{14.004 $\pm$ 2.101}    \\
$a$($R_\odot$)   & \multicolumn{2}{c}{6.300 $\pm$ 0.030}& &\multicolumn{2}{c}{9.179 $\pm$ 0.145}& &\multicolumn{2}{c}{41.225 $\pm$ 1.230}    \\
$r_{pole}$&0.233 $\pm$ 0.006&0.140 $\pm$ 0.005& &0.113 $\pm$ 0.003&0.081 $\pm$ 0.004& &0.056 $\pm$ 0.001&0.087 $\pm$ 0.003\\
$r_{point}$&0.241 $\pm$ 0.007&0.141 $\pm$ 0.005& &0.113 $\pm$ 0.003&0.082 $\pm$ 0.004& &0.056 $\pm$ 0.001&0.090 $\pm$ 0.003\\
$r_{side}$&0.236 $\pm$ 0.006&0.140 $\pm$ 0.005& &0.113 $\pm$ 0.003&0.082 $\pm$ 0.004& &0.056 $\pm$ 0.001&0.089 $\pm$ 0.003\\
$r_{back}$&0.239 $\pm$ 0.007&0.141 $\pm$ 0.005& &0.113 $\pm$ 0.003&0.082 $\pm$ 0.004& &0.056 $\pm$ 0.001&0.089 $\pm$ 0.003\\
$l_3$& \multicolumn{2}{c}{0}& &\multicolumn{2}{c}{0}& &\multicolumn{2}{c}{0} \\
$(\frac{L_{1}}{L_{1}+L_{2}})_V$ &\multicolumn{2}{c}{0.880 $\pm$ 0.008}& &\multicolumn{2}{c}{0.812 $\pm$ 0.080}& &\multicolumn{2}{c}{0.354 $\pm$ 0.011} \\
\hline\noalign{\smallskip}
\end{tabular}}
\tablecomments{0.86\textwidth}{$\Omega_{\rm in}$ is the potential of the inner Roche lobe of the binaries.}
\end{table}

\subsection{Absolute dimensions} \label{sec:four}

\begin{table}[h]
\caption{Absolute parameters of components of three objects.} 
\label{abs} 
\resizebox{\textwidth}{!}{
\centering
\begin{tabular}{ccccccccccccccccc}
\hline\noalign{\smallskip}
Parameter & \multicolumn{2}{c}{ASASSN-V J063123.82+192341.9}
  &  & \multicolumn{2}{c}{ASAS J011416+0426.4} 
  &  & \multicolumn{2}{c}{MW Aur} 
 \\
\cline{2-3}
\cline{5-6}
\cline{8-9}
 & Primary & Secondary && Primary & Secondary & \ & Primary & Secondary\\
 \hline\noalign{\smallskip}
$T_{\rm eff}$(K)&6254 $\pm$ 145&5127 $\pm$ 199& &5621 $\pm$ 58&4883 $\pm$ 66& &7242 $\pm$ 366&6613 $\pm$ 368\\
${\rm log} g$ &4.126 $\pm$ 0.015&4.490 $\pm$ 0.018& &4.378 $\pm$ 0.033&4.564 $\pm$ 0.047& &4.023 $\pm$ 0.048&3.599 $\pm$ 0.054\\
$L$($L_\odot$)&3.062 $\pm$ 0.300&0.485 $\pm$ 0.078& &0.965 $\pm$ 0.073&0.289 $\pm$ 0.032& &13.233 $\pm$ 2.821&23.080 $\pm$ 5.518\\
$M$($M_\odot$)&1.088 $\pm$ 0.016&0.883 $\pm$ 0.016& &0.934 $\pm$ 0.046&0.754 $\pm$ 0.043& &2.052 $\pm$ 0.196&1.939 $\pm$ 0.193\\
$R$($R_\odot$)&1.494 $\pm$ 0.024&0.885 $\pm$ 0.017& &1.035 $\pm$ 0.023&0.751 $\pm$ 0.037& &2.309 $\pm$ 0.078&3.658 $\pm$ 0.161\\
$a$($R_\odot$)   & \multicolumn{2}{c}{6.300 $\pm$ 0.030}& &\multicolumn{2}{c}{9.179 $\pm$ 0.145}& &\multicolumn{2}{c}{41.225 $\pm$ 1.295}    \\
$V$(mag)&13.563 $\pm$ 0.064&18.093 $\pm$ 0.394& &11.264 $\pm$ 0.078&14.274 $\pm$ 0.260&&14.612 $\pm$ 0.509&13.221 $\pm$ 0.292\\

\hline\noalign{\smallskip}
\end{tabular}}
\end{table}

The standard errors of the effective temperatures of secondary components were derived to be 53 K for ASASSN-V J063123.82+192341.9, 30 K for  ASAS J011416 +0426.4 and 34 K for MW Aur (see Table \ref{table:table3}), which are the formal $1\sigma$ errors arising from the WD light curve solution. The corrected standard errors of the effective temperatures of secondary components following $\sqrt{(err_1)^2+(err_2)^2}$, where $err_1$ and $err_2$ are standard errors of the effective temperatures of the primary and the secondary, respectively. The values of the radii of components are obtained according to a relation: $R = ra$, where
$r$ is the mean fractional radius getting from WD code and $a$ is the orbital separation. The mass of the components follows the functions:

\begin{equation}
M_1[M_{\odot}]=1.34068 \times 10^{-2}\frac{1}{1+q}\frac{a^3[R_{\odot}]}{P^2[d]} ,
\end{equation} 

\begin{equation}
M_2[M_{\odot}]= M_1 \cdot q ,
\end{equation}
where $q$ is the mass ratio and $P$ is
the orbital period in days. The individual magnitudes in $V$-band of the stars were derived based on the following equations:

\begin{equation}
V_{1}=V-2.5{\rm log}\frac{1}{1+(L_2/L_1)_V},
\end{equation}
\begin{equation}
V_{2}=V-2.5{\rm log}\frac{(L_2/L_1)_V}{1+(L_2/L_1)_V} ,
\end{equation}
where $V_1$ and $V_2$ are the magnitudes of the primary and secondary
components and $(L_2/L_1)_V$ is the luminosity ratio in $V$ band.
The derived absolute parameters for the three binary systems are presented in Table \ref{abs} .

\section {Discussion and conclusions} \label{sec:five}

In this work, the V-band light curves and RV curves for each eclipsing binary were simultaneously analyzed by using the WD code. It was found that all of the binary systems we studied had a relatively high mass ratio ($\ga 0.8$). It might be a result of observational selection effects, which require the spectral signals of both components presented in a binary spectrum. Meanwhile, the surface potential of each component of our targets is much higher than the potential of their inner Roche lobe. This indicates that the components of these binary systems have not filled the inner Roche lobe and thus they are all well detached eclipsing binaries. In order to estimate the evolutionary status and age of these binary systems, we compared their absolute physical parameters with the theoretical PARSEC stellar evolutionary tracks and isochrones, which produced from the latest version (v1.2S) of PAdova and TRieste Stellar Evolution Code\footnote{\url{http://stev.oapd.inaf.it/cgi-bin/cmd}} \citep[PARSEC;][]{2000A&AS..141..371G, 2012MNRAS.427..127B, 2017ApJ...835...77M}. The best isochrones (age) was selected to minimise the $\chi^2$ function including effective temperature and luminosities of the two components.

\begin{figure}[ht]
\centering
\includegraphics[width=\textwidth]{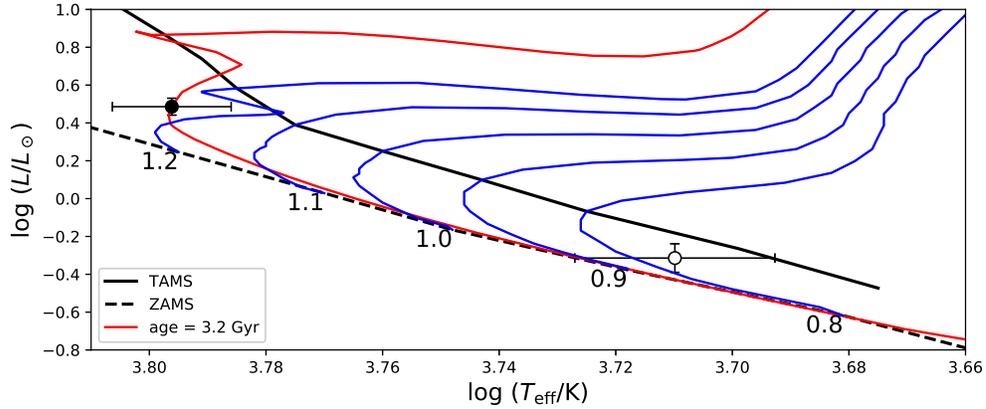}
\caption{Comparison between physical parameters of ASASSN-V J063123.82+192341.9 and PARSEC isochrones in log HR diagram. Filled and open circles represent primary and secondary components, respectively. Red solid line is the isochrone for age = 3.2 Gyr and $Z = 0.0176$, black dashed and solid lines represent Zero-age main-sequence (ZAMS) and terminal-age main-sequence (TAMS), blue lines are the evolutionary tracks were taken from PARSEC models for the stars with different initial masses.
\label{evo1}} 
\end{figure}

\begin{figure}[ht]
\centering
\includegraphics[width=\textwidth]{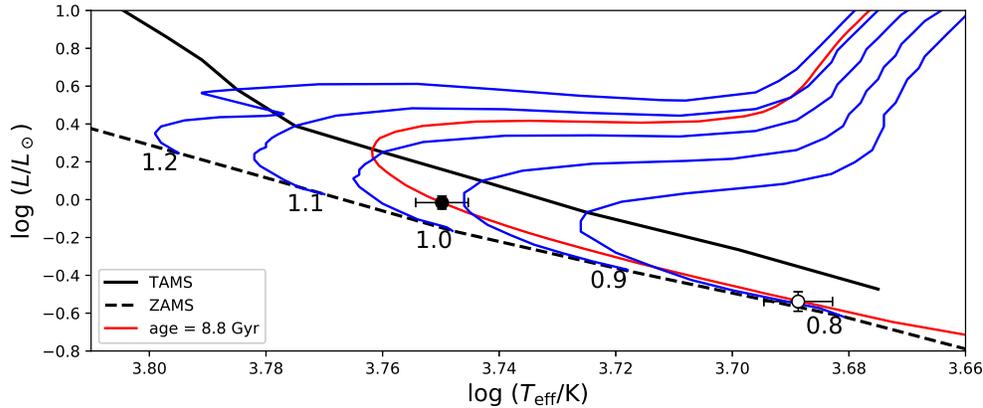}
\caption{Comparison between physical parameters of ASAS J011416+0426.4ASAS J011416+0426.4 and PARSEC isochrones in log HR diagram. Filled and open circles represent primary and secondary components, respectively. Red solid line is the isochrone for age = 8.8 Gyr and $Z=0.0200$, black dashed and solid lines represent ZAMS and TAMS, blue lines are the evolutionary tracks were taken from PARSEC models for stars with different initial masses. 
\label{evo2}} 
\end{figure}

\begin{figure}[ht]
\centering
\includegraphics[width=\textwidth]{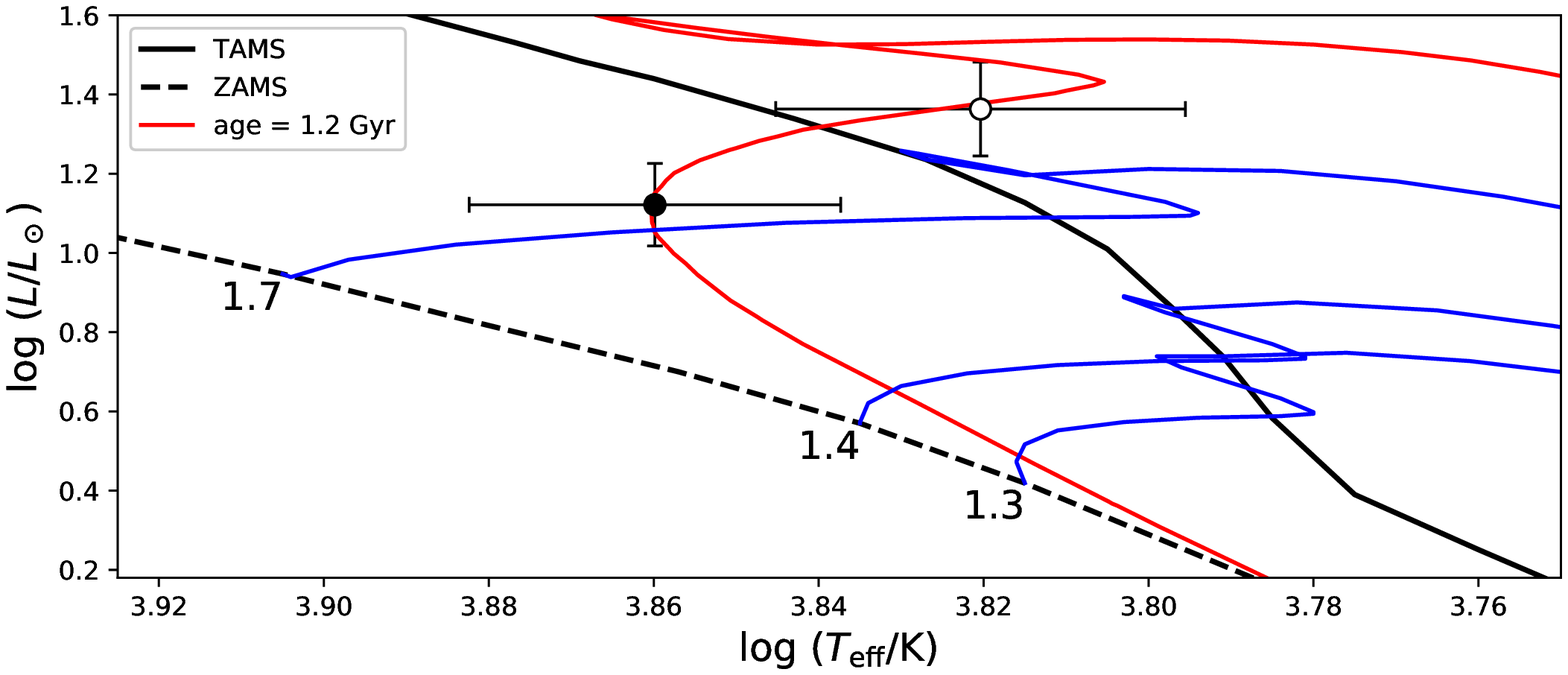}
\caption{Comparison between physical parameters of MW Aur and PARSEC isochrones in log HR diagram. Filled and open circles represent the primary and secondary components, respectively. Red solid line is the isochrone for age = 1.2 Gyr and $Z=0.0199$, black dashed and solid lines represent ZAMS and TAMS, blue lines are the evolutionary tracks were taken from PARSEC models for the stars with different initial masses.
\label{evo3}} 
\end{figure}

Figure~\ref{evo1} shows the locations of two components of  ASASSN-V J063123.82+192341.9 in the Hertzsprung-Russell (HR; ${\rm log}T_{\rm eff}-{\rm log} L$) diagram. It was found that the two components are evolved on the main sequence (MS) stage with an age of about 3.2 Gyr and a metallicity of $Z=0.0176$. According the age and the HR diagram of this binary system, it was inferred that both components in this binary system have not yet undergone the mass transfer and their evolutions should be similar to those of single stars. The same character can be found in Figure~\ref{evo2}, where showed the results of the binary system ASAS J011416 +0426.4. It was found in Figure~\ref{evo2} that both components of ASAS J011416 +0426.4 are also evolved on the MS stage and have an age of 8.8 Gyr and a metallicity of $Z=0.0200$. The results of binary MW Aur are shown in Figure~\ref{evo3}. As seen from Figure~\ref{evo3}, it was found that the  more massive and hotter primary component in this abnormal binary system is evolving on the MS stage, while its secondary component has just evolved away from MS stage, implying that the less massive component seemly evolves more rapidly than its more massive one do. The most likely explanation for this situation might be that some mass had been periodically lost through the Lagrangian point L$_1$ from the present less massive component close to periastron for a binary system with large eccentricity \citep{2005MNRAS.358..544R}. Therefore, the present less massive component shrank and divorced from its inner Roche lobe, and thus the mass transfer is stopped and this binary show a detached configuration at present.  Finally, the age of MW Aur was estimated as 1.2 Gyr and a metallicity of $Z=0.0199$. If the mass transfer had indeed been taken place in this binary, and thus the age and metallicity estimates for this object should be inaccurate, we will use the Modules for Experiments in Stellar Astrophysics (MESA) evolution code \citep{2011ApJS..192....3P, 2013ApJS..208....4P} to investigate the evolutionary status of MW Aur in detailed in our future work. 

\begin{figure}[ht]
\centering
\includegraphics[width=\textwidth]{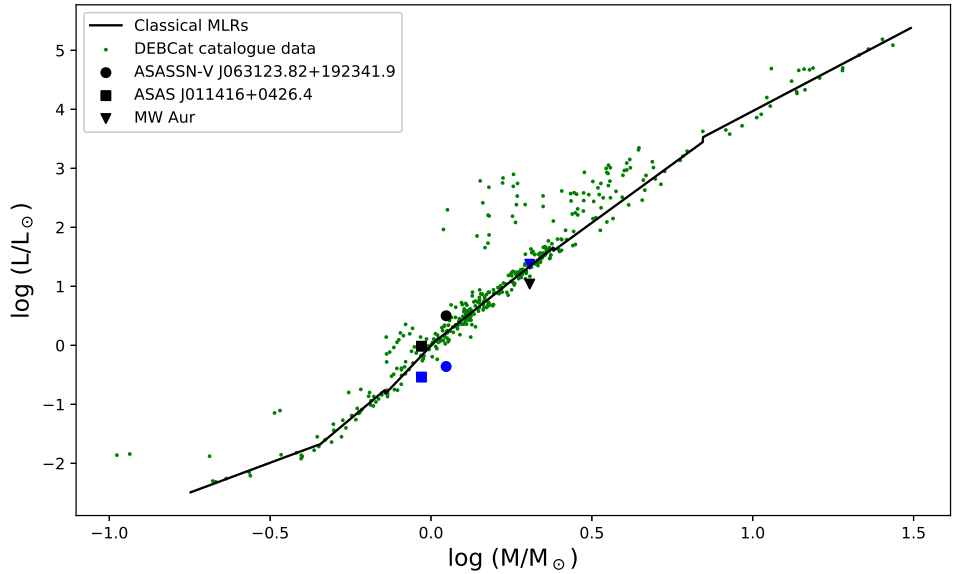}
\caption{Comparison between the derived parameters and the classical MLR. The black line is the MLR, black and blue symbols represent primary and secondary components, green points represent the data from  DEBCat catalogue \citep{2015ASPC..496..164S}.
\label{M_l}} 
\end{figure}

We found that the best fitted metallicity are consistent with those shown in the Table~\ref{table:table1} and the studied systems are all have a solar metallicity. We also compared our results with the classical MS mass-luminosity relation \citep[MLR, ][]{2018MNRAS.479.5491E}, the results are shown in Figure~\ref{M_l}. It was found in Figure~\ref{M_l} that the three eclipsing binaries are evolving on or near the MS stage, which correspond to the result of the above analysis.

The orbital and physical parameters for three well-detached eclipsing
binaries were derived from their light curves and RV curves based on WASP, ASAS-SN, ASAS and LAMOST-MRS data. The accurate parameters are very important for testing the stellar structure and evolution models \citep{2009ApJ...700.1349T}.
LAMOST-MRS survey lasts for five years,  it will find more and more double-lined spectroscopic binaries and play an important role in determining the accurate parameters of both components of eclipsing binaries. We will study other eclipsing  binaries found from the five-year LAMOST-MRS survey and determine their absolute parameters in the future.

\begin{acknowledgements}
We thank the anonymous referee for the helpful comments and suggestions which improvement this work greatly. This work was supported by the National Natural Science Foundation of China (NSFC)
under No. 11773065.

\end{acknowledgements}
\bibliography{bib}{}
\bibliographystyle{aasjournal}
\appendix                  

\end{document}